# Synthesis and structure of high-quality films of copper polyphthalocyanine – 2D conductive polymer

Darya M. Sedlovets,[†] Maksim V. Shuvalov,[‡] Yury V. Vishnevskiy,[§] Vladimir T. Volkov,[†] Igor I. Khodos,[†] Oleg V. Trofimov,[†] Vitaly I. Korepanov[†]*

[†] Institute of microelectronics technology and high purity materials RAS, Institutskaya street, 6, Chernogolovka, 142432, Russia; [‡] Chemistry department, Moscow State University 119991, Moscow, Russia; [§] Universität Bielefeld, Lehrstuhl für Anorganische Chemie und Strukturchemie, Universitätsstrasse 25, D-33615, Bielefeld, Germany

**ABSTRACT:** We propose an experimental approach, by which thin films of copper polyphthalocyanine (CuPPC) can be directly synthesized in a chemical vapor deposition (CVD) set-up at mild temperature (420 °C). High polymerization degree and high crystallinity of the films was confirmed by TEM, FTIR and UV-VIS studies; the stacking structure of CuPPC layers was determined, inter-layer spacing was estimated from XRD and TEM electron-diffraction. Quantum-chemical study performed providing support for experimental structure determination and yielding information on the electronic structure.

## 1. Introduction

Two-dimensional electronic conjugation makes metal polyphthalocyanines (PPCs) (fig. 1) a unique class of elementoorganic semiconductors which are of especial interest for science and technology [1–5]. Although such compounds are known for more than 50 years [6,7], their application has been hindered by the lack of processability: PPCs are practically insoluble in all solvents and cannot be melted or evaporated [2]. For these reasons PPCs cannot be treated by the conventional processing methods like spin-coating, sputtering or thermal evaporation. However, in order to access many promising applications of PPCs, it is desirable to obtain them in a form of thin film materials on any arbitrary wafer, in particular on dielectrics.

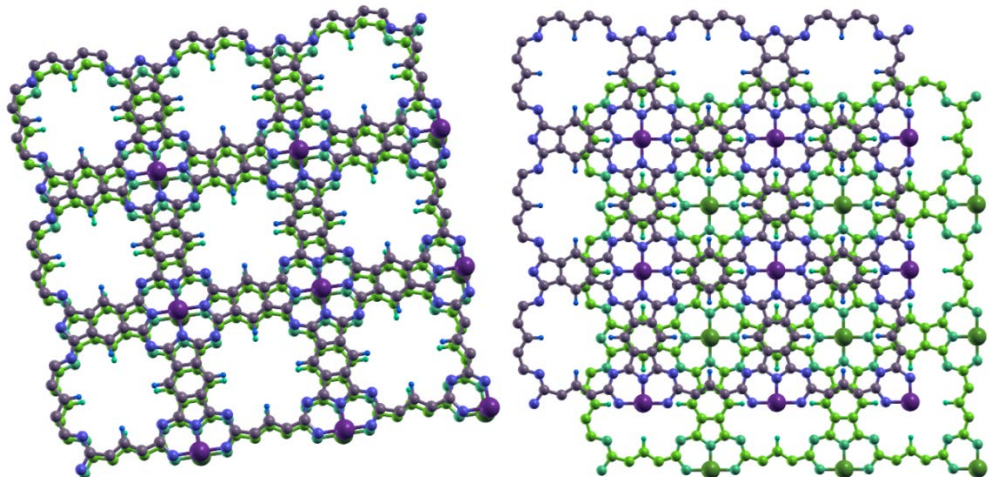

Figure 1. Molecular structure of CuPPC with AA (left) and AB (right) stacking.

The reported synthetic routes for PPCs are based on reaction of metals [8] or metal salts [7] with pyromellitic acid tetranitrile (PMTN) (fig. 2). Recently, the chemical structure of the polymer formed by this reaction was confirmed by the scanning tunneling microscopy (STM) for the case of Fe [5]. However, only nano-scale samples were achieved, while the experimental set-up was quite complex (ultra-high-vacuum conditions and atomically clean surfaces). Other previously reported synthetic approaches include: (i) the reaction of pre-sputtered copper (1.5-30 nm layers) with PMTN vapor at T > 350°C [8] in sealed ampoules; (ii) double-source evaporation of copper and PMTN [9] with subsequent annealing. The significant drawbacks of the first approach are the complexity of experimental set-up and the inevitable oxidation of copper surface when exposed to air even for a short time (which is especially important for the thin films of metal). In a double-source evaporation (ii) the film thickness was not precisely controlled, so the resulting films were quite thick (~1 μm), which is also undesirable. Based on the reported IR spectra of PPC materials, it can be concluded that PPC materials synthesized by all these methods had significant amounts of low-molecular weight fractions and poor structural uniformity [9,10].

In this work we propose a different experimental approach to the synthesis of CuPPC thin films, namely the reaction of PMTN with copper in a CVD set-up. We prove that in this way the conductive CuPPC films of high uniformity can be obtained. We use infrared spectroscopy, transmission electron microscopy, XRD, UV-visible absorption and sheet resistance measurements to prove the structure and investigate the properties of the films. The quantum-chemical calculations provide support for the interpretation of the X-ray and TEM electron diffraction pattern. In discussion of FTIR spectra and electronic structure, we compare CuPPC with copper phthalocyanine (CuPC), referring to it as a monomer.

One should understand however, that this analogy is of limited applicability, because CuPC has 4 benzene rings per molecule, while CuPPC has only 2 of them per unit cell. But since the symmetry of the former and the latter is the same, this comparison is used hereafter for the reason of convenience.

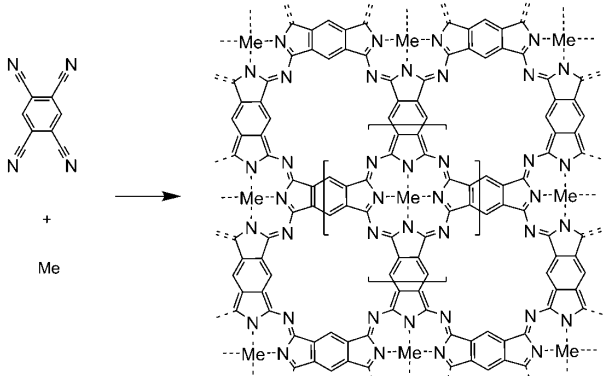

Figure 2. Synthesis of polyphthalocyanines from metal and pyromellitic acid tetranitrile.

The fact that for over 50 years the crystal structure of CuPPC has not been understood is also a result of the poor processability: it is a very challenging experimental task to obtain high-quality material suitable for XRD measurements. Moreover, once the XRD pattern is obtained, its interpretation is also not straightforward; in the present work it would be impossible to assign the XRD peaks without the help of quantum-chemical studies.

Electronic properties of CuPPC were also a matter of discussions [7,9,11,12]. Until now it was even unclear if CuPPC is semiconductor or semimetal. On one hand, the reliable experimental data could not been achieved before, since the previously obtained materials always contained high concentration of monomeric and oligomeric units. There are only estimations of charge transport activation energy (0.1-0.2 eV [7,9]). Also, it has been shown that CuPPC has a high dielectric constant (>10,000), which makes it an attractive material for electro-active composites [3]. On the other hand, to the best of our knowledge, nobody attempted theoretical study of the bulk polymer electronic structure. In fact, such theoretical study would not have any practical value without understanding of the polymer structure, especially the layer stacking mode. The only relevant study by now is the band gap estimation for monolayer made by Zhou [11], suggesting a value of 0.31 eV.

In this work we use the combination of quantum-chemical calculations and experiment to assign the crystal structure and predict electronic properties of CuPPC. Computational results provide a good support for the interpretation of experimental data, and both approaches give a consistent description of the structure.

## 3. Results and discussion

### 2.1. Chemical reaction and structure of the films

After the reaction of copper with PMTN vapors (fig. 2), notable changes were observed. In most of experiments with the foil the copper changed color to purple. By the AFM measurements of the film transferred to $SiO_2$/Si substrate it was found that the resulting film has a thickness of >100 nm. In the experiments with pre-sputtered metal, color changed to yellow or brown (observed on quartz substrate) for more than 1 hr experiments, while in short experiments (0.5 hr) it still contained some amount of red tint due to incompletely reacted copper. It was found that the red color disappears after 1 hr for 8 nm copper layer, so we may safely assume that for longer time (>2 hr) and thinner layer (4 nm) the reaction is complete. A 10 nm film of CuPPC was formed in this case, which can be easily observed in optical microscope on $SiO_2$/Si wafer (fig. 3).

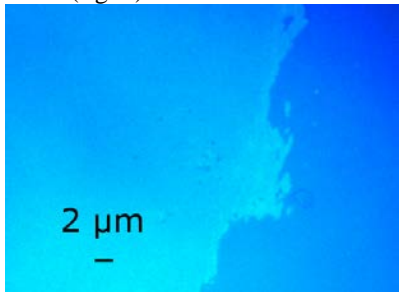

Figure 3. Optical microscope image of CuPPC film near the edge. Left: CuPPC, right: $SiO_2$/Si wafer. Contrast adjusted for better visibility.

Essential information on the molecular structure of CuPPC is provided by FTIR spectra. Figure 4 shows the spectrum of CuPPC material obtained in this work by CVD at 420°C as compared to the spectrum of CuPC. As expected, there is no one-to-one correspondence between polymer and monomer, taking into account the difference in their electronic structures (see below) and collectivization of vibrational modes in the spectrum of polymer.

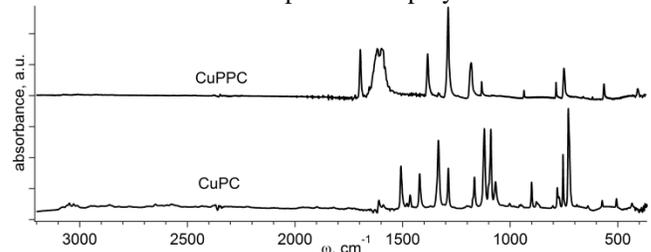

Figure 4. FTIR spectra of CuPPC (top curve) as compared to copper phthalocyanine (CuPC, bottom curve)

No features in the 2900-3100 $cm^{-1}$ region were observed in the spectrum of polymer, nor distinct signal from the end groups (C≡N stretching mode in case of low polymerization degree should be observed at around 2220 $cm^{-1}$ [9]). There is a general agreement between the spectrum of CuPPC obtained in this work and the published spectra [9,10]. The main difference is that the bands for the CVD material are much sharper (~15 $cm^{-1}$ in our work as compared to >100 $cm^{-1}$ for previous reports). Thus, we conclude that the polymerization degree and material uniformity in our case are much higher than in any of the previous reports on CuPPC.

From the comparison of CuPC and CuPPC spectral patterns (fig. 4) and the known information on the CuPC vibrations [13] it is possible to suggest a tentative assignment for the CuPPC spectrum. Consequently, we can assume that the intense features at 1618 and 1596 cm$^{-1}$ belong to CC stretching modes together with the CH in-plane bendings (IPB) of PC; 1383 and 1288 cm$^{-1}$ bands probably correspond to the CN stretching of the pyrrole ring. It is likely that the 1182 and 1133 cm$^{-1}$ bands have a strong contribution of CH IPB, which is also confirmed by the low intensity of these bands in CuPPC as compared to CuPC (CuPPC has 4 hydrogen atoms per tetrapyrrole ring as compared to 16 for CuPC). The 936 cm$^{-1}$ band might correspond to CN stretching and CNC bending vibrations of the PC ring; 787, 750 and 564 cm$^{-1}$ bands of CuPPC probably correspond to 753, 726 and 571 cm$^{-1}$ bands of CuPC, which are attributed to isoindole ring deformations, while the 408 cm$^{-1}$ band corresponds to CCC out-of-plane bendings (benzene ring), which give the 433 cm$^{-1}$ band in CuPC spectrum.

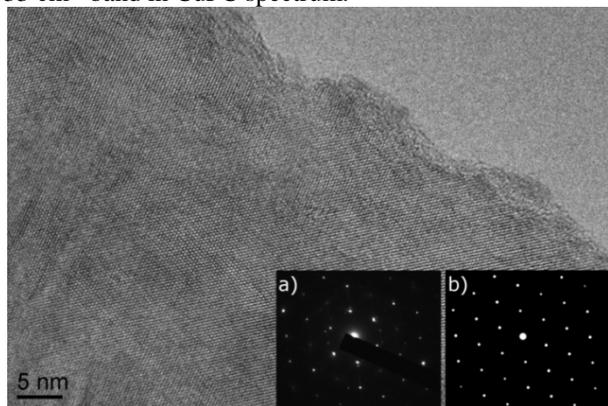

Figure 5. TEM image taken from CuPPC film. Inset: ED images: experimental (a) and generated from the ab initio geometry (b)

TEM study shows that the CVD film has a polycrystalline structure with the characteristic crystallite size of at least several tens of nanometers (fig. 5). From the comparison of experimental and simulated electron diffraction (ED) patterns we can make important conclusions on the structure of the polymer (fig. 5). First, only the AA layer stacking (fig. 1) gives a good match between experimental and simulated ED patterns. This result is corroborated by quantum-chemical calculations which show energetic preference for the AA stacking: the energy difference between AA and AB structures per unit cell is ~700 cm$^{-1}$. The interlayer distance measured from TEM image is 0.34 nm.

The XRD pattern taken from a relatively thick film grown on Cu foil (experiment time: 8 hr) is shown on fig. 6. Simulation of the diffraction pattern based on the quantum-chemistry optimized structure shows a good agreement with the experimental data (cell parameters were slightly adjusted to match experimental data: 0.319x1.053 nm). The interlayer distance derived from XRD is 0.319 nm, which is slightly smaller that observed for the monomer (0.34 nm [14]). Thus, we have three estimations for interlayer distance: 0.338 nm from quantum-chemical calculations, 0.34 nm from TEM image and 0.319 nm from XRD. The latter value, probably, should be considered as the most reliable.

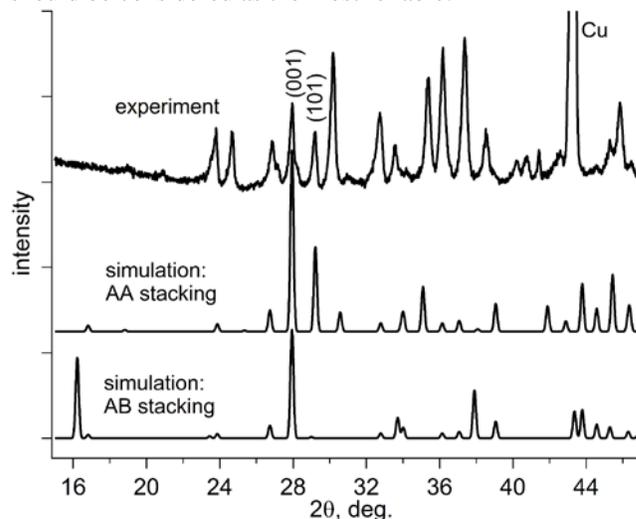

Figure 6. XRD pattern of CuPPC. Top: experimental curve, middle: simulated for AA stacking, bottom: simulated for AB stacking.

### 2.2. Optical properties and electronic structure

Optical measurements show that absorption of the CuPPC films (fig. 7) continues far beyond the monomeric CuPC absorption threshold (~700 nm), which is a clear proof of the extended conjugated π-electronic system. Another important observation is the absence of the ~740 nm absorption band, which was reported for this polymer previously [10]. Note that in the cited work the measurements were made in the solution of concentrated H$_2$SO$_4$, which implies the low-molecular weight fraction of the polymer, while the material obtained in the present work was found to be completely insoluble in sulfuric acid. The latter is an additional confirmation of high polymerization degree of the CuPPC obtained in the present work.

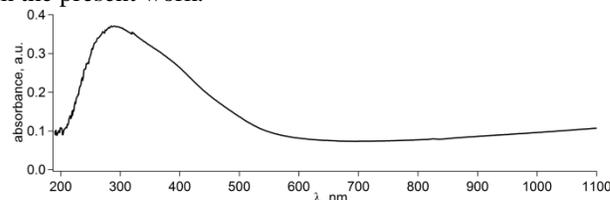

Figure 7. Absorption spectrum of the CuPPC thin film (10 nm on quartz plate).

Since the ~700 nm band is attributed to the lowest-energy electronic transition of the CuPC, and in the CuPPC both HOMO and LUMO changed shape and energies as compared to CuPC, the absence of 740 nm band for the polymer provides more evidence of a high polymerization degree.

Recent calculations made for the CuPPC monolayer by Zhou et al. [11] (DFT with the PBE functional) showed that the energy difference between antiferromagnetic and ferromagnetic states of CuPPC is ~16 cm$^{-1}$ per unit cell in favor of the ferromagnetic one. This difference is quite small as compared to the value for MnPPC, which is 254 cm$^{-1}$. This implies that CuPPC should have a paramagnetic character, being similar to monomeric CuPC (see experimental results on the latter in the work [15]), and that the spin interaction between the neighboring cells is relatively weak.

DFT calculations (fig. 8) show that the highest occupied crystal orbital (HOCO) in α-channel (α has 1 electron more than β) is partially filled. This orbital seems to be originated from Cu atomic orbitals (fig. 10). Both HOCO-1 and lowest unoccupied (LUCO) come from the π-orbitals of phthalocyanine, they are both delocalized. Such band structure suggests that CuPPC should possess semimetallic conductivity. Calculated density of states (DOS) shows low values near Fermi level (fig. 9); moreover, the peaks located c.a. at ±0.35 eV belong to orbitals with different spin, which should make charge transport even more difficult. Analysis shows that these 2 orbitals come from Cu $d$-orbitals (see fig. 10, center).

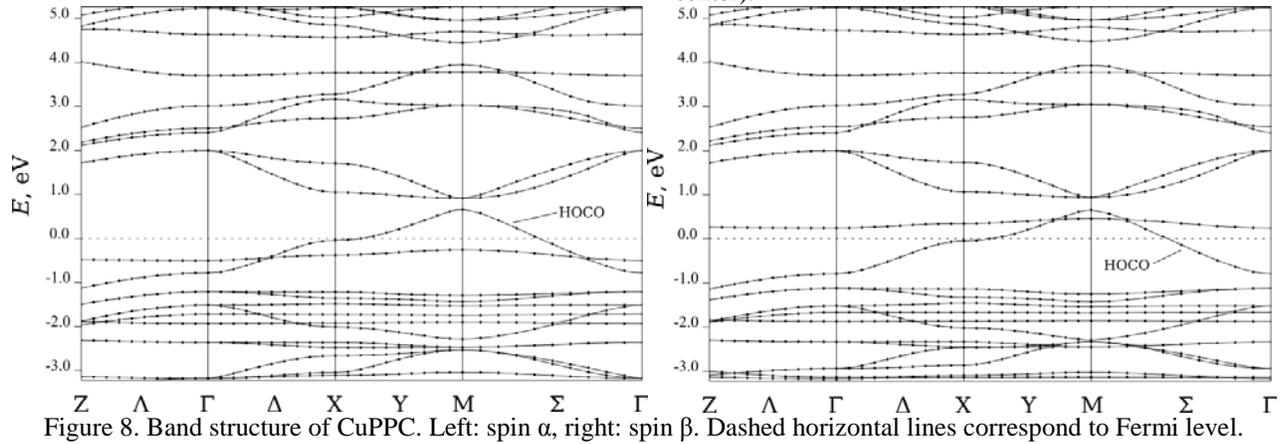

Figure 8. Band structure of CuPPC. Left: spin α, right: spin β. Dashed horizontal lines correspond to Fermi level.

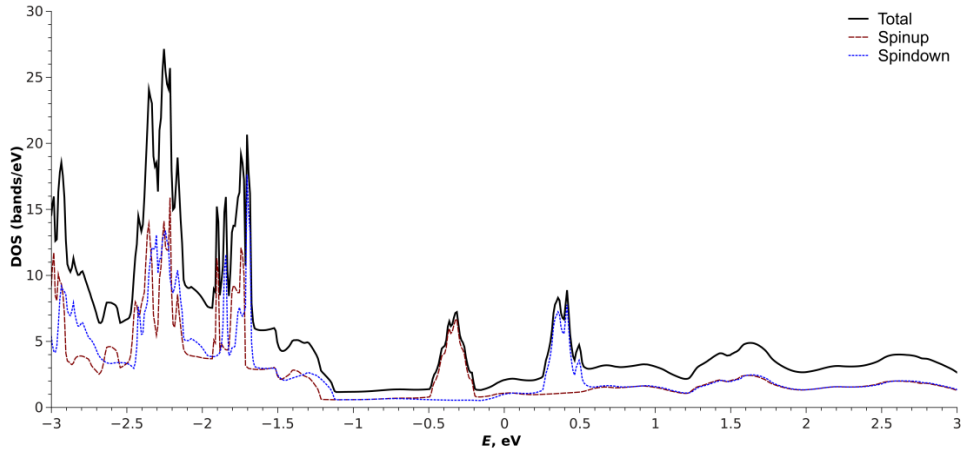

Figure 9. Calculated density of states for CuPPC: total (black curve), α channel (red curve), β channel (blue curve). Peaks at ±0.35 eV belong to orbitals, coming from Cu $d$-orbitals (see fig. 10, center).

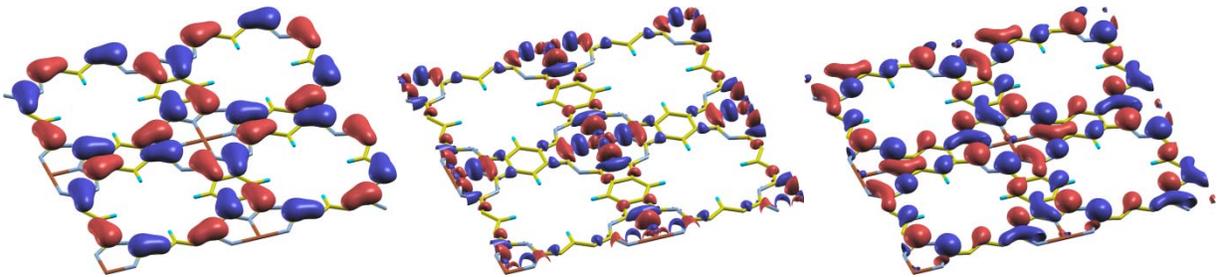

Figure 10. Frontier crystal orbitals of CuPPC. Left: HOCO-1, middle: HOCO (partially filled), right: LUCO.

Sheet resistance of 10 nm film of CuPPC was measured to be 12 kOhm/sq under ambient conditions and 9 kOhm/sq after annealing at 130 °C in argon atmosphere (the resistance of the original 4 nm Cu film was around 80 Ohm/sq). Thus, the conductivity of this film is ~10 times lower than that of highly uniform graphene [16]. Taking into account remarkably high electron mobility of graphene and relatively high amount of grain boundaries in CuPPC, this value is also consistent with quantum-chemical results. Resistance of the 10 nm CuPPC film does not exhibit strong change with the temperature (fig. 11); below 423 K the resistance change is completely reversible, and at higher temperatures the film undergoes some changes of either mechanical of chemical nature, which result in the decrease of conductivity.

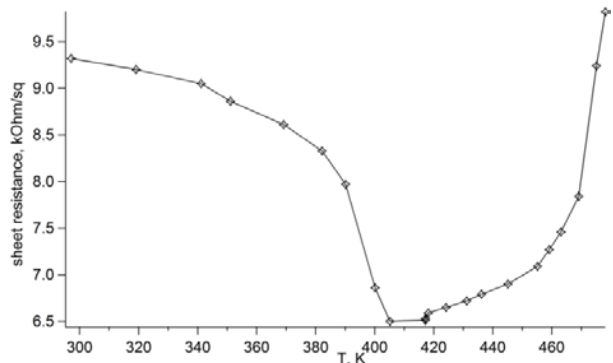

Figure 11. Temperature-dependent sheet resistance of 10 nm CuPPC film

The estimation of the charge transport activation energy from (1/T) vs. log(s) dependence in 300-400 K range gives the value of ~$1*10^{-5}$ eV, which is 4 orders of magnitude less than that measured previously for bulk polymer [7,9]. This temperature behavior is in a better correspondence with what one should expect from semi-metal. Such different result can be easily explained by our quantum-chemical calculations: the material with higher polymerization degree (this work) should have lower band gap (CuPC has a gap of ~1.5 eV [17], and this value decreases with increasing elec-tronic conjugation). On the other hand, the direct comparison between the bulk material and the thin film is of limited applicability, because in the latter case we should also consider surface effects. The detailed experimental study of electronic properties of CuPPC should be a subject of extensive future research.

**2.3. Processability**

Since the PMTN vapor pressure is controlled by the temperature at the evaporation zone, which was kept constant in all experiments, the film thickness can be controlled by the experiment time (in case of metal foil) or by the thickness of the metal layer (in case of pre-sputtered metal). In the latter case, it is easy to obtain PPC films of any desired geometry, for example by using deposition masks at a stage of metal pre-sputtering.

In experiments with metal foil, the PPC films were transferred onto dielectric ($SiO_2$/Si) for conductivity measurements. It was found, however, that the films transferred in this way have no electric conductivity. By optical and electronic microscopy was found that they contain many microscopic defects like cracks and wrinkles (this is in contrast to graphene, which was transferred successfully in our group). The reason for this is most likely the thickness: on the foil, thick film of ~100 nm is formed, which results in a poor flexibility and hence easy damage of the film.

## 3. Conclusions

In this work we developed an approach to the direct synthesis of thin films of CuPPC. Although the material was obtained by a well-known chemical reaction, in the present work we achieved the polymerization degree and material uniformity much higher than reported before. This conclusion is supported by FTIR, UV-vis spectra and by the complete insolubility of the films in sulfuric acid. TEM, XRD and quantum-chemical studies gave consistent data on the structure of the polymer, including the stacking mode (AA is preferred) and inter-layer distance (0.32 nm). Thermally activated mechanism for charge transport, previously reported for low-polymerization degree material, was not observed in this work. Quantum-chemical calculations predicted that CuPPC should have semimetallic conductivity, since the HOCO (originated from Cu orbitals) is partially filled, and this finding provides a good explanation of the experimental temperature dependence of sheet resistance.

The synthetic technique proposed here solves the problem of otherwise poor processability of CuPPC, and this is an important step for the chemistry and technology of 2D polymers. The deposition technique does not require complex experimental set-up; the CuPPC film can be easily prepared on a wide variety of substrates, including dielectrics.

## 4. Methods

**4.1. Reaction of PMTN with copper and transfer to dielectric wafers**

Cu films on dielectric ($SiO_2$/Si, quartz or KBr) were prepared by diode radio-frequency sputtering of Cu targets on a Z-400 installation [18]. Prior to deposition, the working chamber was evacuated to $1*10^{-4}$ Pa. The rate Cu deposition was 0.34 nm/s. For experiments with foil, freshly electropolished Cu foil was used.

Reactions in a CVD set-up were performed in a quartz tube reactor, placed in a two-zone oven. During the reaction, the pressure of $1*10^3$ Pa was maintained. PMTN was evaporated in a first zone, while the wafers were placed in a second one. The temperatures in the two zones were controlled by thermocouples. Reaction temperature was 420°C; the metal films were annealed in hydrogen atmosphere for 15-30 min prior to deposition to remove any oxides from the surface. Reaction time was 0.5-8 hr. Temperature in the first zone was adjusted to make the PMTN evaporation rate to be 8 mg/hr.

In CVD experiments with pre-sputtered copper, KBr was used as a substrate for FTIR measurements. Quartz glass was used for optical measurements. For electrical measurements samples were obtained on the thermally oxidized silicon (300 nm $SiO_2$).

In experiments with copper foil, the graphene technology procedure [19] was applied to transfer synthesized films from the foil to dielectric. As the first step, the foil was coated with polymethylmetacrylate (PMMA). Next, the metal was dissolved in $FeCl_3$ solution and the PPC film was transferred on the arbitrary surface with subsequent dissolution of PMMA in organic solvents.

**4.2. Analytical characterization**

The sheet resistance, being the conventional measure for ultrathin graphitic films, was chosen as a basic characteristic of conductivity [16]. Electrical measurements were made with square samples of at least 3x3 mm size.

Optical measurements were made with a Specord-50 spectrometer. TEM studies were performed on a JEOL JEM 2000FX transmission electron microscope. FTIR spectra were taken with a Bruker IFS-113v spectrometer in the 400-4000 cm$^{-1}$ range under 1 cm$^{-1}$ resolution. XRD was taken from the sample of CuPPC on SiO$_2$/Si wafer with Cu K$\alpha$ irradiation using the Bruker D8 Advance X-ray diffractometer in grazing incidence configuration.

### 4.3. Computational procedures

Calculation for 3D periodic structure were performed using Quantum Espresso package with DFT (PBE), plane wave basis set (ultrasoft Vanderbilt pseudopotential, $E_{cut}$ = 35 Ry; $E_{rho}$ = 420 Ry) and 8x8x8 $k$-points per translational unit with uniform automatic Monkhorst-Pack grid.

For fitting the simulated TEM electron diffraction pattern to the experimental one the SingleCrystal software was used; ab initio (DFT) geometry parameters were taken to generate theoretical TEM pattern.

XRD pattern was generated with a Mercury software package. Ab initio geometry was taken as a starting point; both AA and AB stacking options were compared with the experimental pattern. Cell parameters were slightly adjusted to match the experiment.

## 5. Acknowledgements

We acknowledge Dr. D.V. Irzhak (IMT RAS) and Dr. I.S. Bushmarinov (INEOS RAS) for XRD measurements, Dr. Yu.N. Khanin (IMT RAS) for discussions, Dr. N.A. Yarykin (IMT RAS) and Dr. V.M. Senyavin (MSU) for spectral measurements. Yu.V. thanks for support the Alexander von Humboldt Foundation.## 6. References

[1] X. Feng, L. Liu, Y. Honsho, A. Saeki, S. Seki, S. Irle, Y. Dong, A. Nagai, D. Jiang, Angew. Chem., Int. Ed. 51 (2012) 2618–2622.
[2] D. Wöhrle, Macromol. Rapid Commun. 22 (2001) 68–97.
[3] Q.M. Zhang, H. Li, M. Poh, F. Xia, Z.Y. Cheng, H. Xu, C. Huang, Nature 419 (2002) 284–287.
[4] J. Sakamoto, J. van Heijst, O. Lukin, A.D. Schlüter, Angew. Chem., Int. Ed. 48 (2009) 1030–1069.
[5] M. Abel, S. Clair, O. Ourdjini, M. Mossoyan, L. Porte, J. Am. Chem. Soc. 133 (2011) 1203–1205.
[6] W.C. Drinkard, J.C. Bailar, J. Am. Chem. Soc. 81 (1959) 4795–4797.
[7] A. Epstein, B.S. Wildi, J. Chem. Phys. 32 (1960) 324–329.
[8] D. Wöhrle, V. Schmidt, B. Schumann, A. Yamada, K. Shigehara, Ber. Bunsenges. Phys. Chem. 91 (1987) 975–981.
[9] M. Yudasaka, K. Nakanishi, T. Hara, M. Tanaka, S. Kurita, M. Kawai, Synth. Met. 19 (1987) 775–780.
[10] D. Wöhrle, U. Marose, R. Knoop, Die Makromolekulare Chemie 186 (1985) 2209–2228.
[11] J. Zhou, Q. Sun, J. Am. Chem. Soc. 133 (2011) 15113–15119.
[12] P. Gomez-Romero, Y.S. Lee, M. Kertesz, Inorg. Chem. 27 (1988) 3672–3675.
[13] Z. Liu, X. Zhang, Y. Zhang, J. Jiang, Spectrochim. Acta, Part A 67 (2007) 1232–46.
[14] N.S. Lebedeva, E.A. Mal'kova, A.I. V'yugin, Review Journal of Chemistry 2 (2012) 20–50.
[15] B.N. Achar, K.S. Lokesh, J. Solid State Chem. 177 (2004) 1987–1993.
[16] J.W. Suk, A. Kitt, C.W. Magnuson, Y. Hao, S. Ahmed, J. An, A.K. Swan, B.B. Goldberg, R.S. Ruoff, ACS Nano 5 (2011) 6916–6924.
[17] N. Marom, O. Hod, G.E. Scuseria, L. Kronik, J. Chem. Phys. 128 (2008) 164107.
[18] V. Volkov, V. Levashov, V. Matveev, L. Matveeva, I. Khodos, Y. Kasumov, Thin Solid Films 519 (2011) 4329–4333.
[19] X. Li, W. Cai, J. An, S. Kim, J. Nah, D. Yang, R. Piner, A. Velamakanni, I. Jung, E. Tutuc, S.K. Banerjee, L. Colombo, R.S. Ruoff, Science 324 (2009) 1312–1314.